\documentclass[two column, prA]{revtex4}%
\usepackage{amsmath}
\usepackage{graphicx}
\usepackage{amsfonts}
\usepackage{amssymb}%
\providecommand{\U}[1]{\protect\rule{.1in}{.1in}}

\begin{document}
\title{Photon counting by inertial and accelerated detectors}
\author{Margaret Hawton}
\email{margaret.hawton@lakeheadu.ca}
\affiliation{Department of Physics, Lakehead University, Thunder Bay, ON, Canada, P7B 5E1}

\begin{abstract}
Bases of exactly localized Minkowski and Rindler states on spacelike
hypersurfaces are used to describe inertial and accelerated photon counting
devices. It is found that the spacetime coordinates of photons absorbed by a
pair of counteraccelerating detectors in causally disconnected Rindler wedges
are correlated. If a photon is absorbed by a single accelerated detector the
Minkowski vacuum collapses to a state containing at least one photon and that
photon can be absorbed by an inertial detector.

\end{abstract}
\maketitle

\section{Introduction}

Absorption in a small photon counting detector provides a measurement of
position with an accuracy determined by the size of the device. For quantum
states containing a small number of photons an array detector counts the
photons incident on its surface and records their locations in spacetime. The
position measurement performed by a photon counting array detector can be
described by a positive operator valued measure (POVM) whose elements are
projectors onto exactly localized states
\cite{Tsang,HawtonPOVM,HawtonPhotonLocation}. However, these localized states
have paradoxial properties; for example any field describing a localized state
is itself nonlocal \cite{BB96}. The resolution of these paradoxes will be
discussed in Section VII but for nonlocality it is roughly as follows: To
count photons a detector should be thick enough to absorb all photons incident
on its surface. The probability for an individual atom to absorb a photon is
proportional to the absolute square of the electric field which is
proportional to frequency, but penetration depth is inversely proportional to
frequency and these factors cancel leaving number density.

The theoretical description of particle absorption and emission is based on
quantum field theory (QFT) in which positive frequency modes are associated
with annihilation and negative frequencies with creation. Each application of
a creation operator adds a particle to the field and all inertial observers
will agree on the number of particles present. The vacuum is the zero particle
state and an inertial detector absorbs no particles from the Minkowski vacuum.
Beyond the realm of inertial detectors in flat spacetime the separation of the
field modes into positive and negative frequencies and the definition of the
vacuum state are not unique \cite{BirrellDavies}. This observer dependence of
the particle content of field theory is of fundamental importance and leads to
creation of Hawking particles near a black hole \cite{BlackHole} and
absorption of particles from the vacuum by an accelerated detector. The latter
phenomenon, known as the Unruh effect \cite{Unruh}, will be discussed here.

The Unruh-deWitt detector commonly used to model an accelerated device is a
two-level point monopole coupled to a real massless scalar field that can only
absorb particles in a very narrow band of frequencies \cite{Unruh,MMdelRay}.
The photon counting detector considered here utilizes a semiconductor band
structure to allow absorption of a wide band of frequencies. Biasing of the
semiconductor pn-junction separates any electron-hole created so that the
photon is not reemitted. It counts photons in the sense that any photon
crossing its surface will eventually be absorbed. Photon counting detectors
need not be small unless high spatial resolution is required. Instead a pixel
must be large if it is to absorb the low frequency photons that characterize
the Unruh effect. The function of the basis of exactly localized states is to
calculate photon probability density as a function of spacetime location on a hypersurface.

In this paper the family of POVMs describing photon counting array detectors
proposed in \cite{Tsang,HawtonPOVM,HawtonPhotonLocation} is extended to
include uniformly accelerated devices described in flat spacetime by Rindler
coordinates \cite{Rindler}. It was proved in \cite{HawtonPOVM} that the
probability density for absorption equals the absolute square of the
projection of the photon state vector onto the localized states. For inertial
detectors this was generalized to a covariant formalism in
\cite{HawtonPhotonLocation}. Here both Minkowski and Rindler localized states
are defined on spacelike hypersurfaces and a transformation between the
Rindler and Minkowski localized bases is derived. The formalism is applied to
the Unruh effect.

The plan of the paper is as follows: Section II defines the Rindler
coordinates used to describe accelerated detectors and Section III describes
photon counting experiments. Section IV is concerned with the 4D photon number
density operator and the 2D invariant photon scalar product. Section V
summarizes the properties of the standard Minkowski and Rindler plane waves
and provides a new derivation of the 2D Bogoliubov coefficients. In Section VI
the Minkowski and Rindler localized states are defined and the transformation
coefficients between them are derived. In Section VII the paradoxical
properties of localized staetes are discussed. In Section VIII the localized
state formalism is applied to the Unruh effect and in Section IX we conclude.
Natural units in which $\hbar=c=1$ are used. In 4D with metric signature
$\left(  -,+,+,+\right)  $ they are $x^{\mu}=\left(  t,\mathbf{x}\right)
=\left(  t,x,y,z\right)  $ and $k^{\mu}=\left(  \omega,\mathbf{k}\right)  .$
In 2D the Minkowski variables are $\left(  t,x\right)  $ and $\left(
\omega,k\right)  $ and the Rindler variables are $\left(  \eta,\xi\right)  $
and $\left(  \Omega,K\right)  $.

\section{Rindler coordinates}

Uniformly accelerated detectors are described by the Rindler coordinates
$\eta$ and $\xi$. In wedge I $\left\vert t\right\vert <x$ and $\eta$ and $\xi$
are defined by \cite{BirrellDavies,Carroll}
\begin{align}
t  &  =a^{-1}\exp\left(  a\xi\right)  \sinh\left(  a\eta\right)
,\ \label{RinderCoordinates}\\
x  &  =a^{-1}\exp\left(  a\xi\right)  \cosh\left(  a\eta\right)  ,\nonumber
\end{align}
where $a$ is a positive constant. In wedge II where $\left\vert t\right\vert
<-x$
\begin{align}
t  &  =-a^{-1}\exp\left(  a\xi\right)  \sinh\left(  a\eta\right)
,\ \label{RindlerCoordinatesII}\\
x  &  =-a^{-1}\exp\left(  a\xi\right)  \cosh\left(  a\eta\right)  .\nonumber
\end{align}
Eqs. (\ref{RinderCoordinates}) and (\ref{RindlerCoordinatesII}) can be
inverted and combined to give%
\begin{equation}
a\left(  \xi-\epsilon\eta\right)  =\ln\left(  a\left\vert x-\epsilon
t\right\vert \right)  \label{InOut}%
\end{equation}
where $\epsilon$ labels the propagation direction \cite{Fuentes}.

On the hypersurface $\eta=\eta^{\prime}$ all elements of the POVM have a
common velocity $\beta=dx/dt=\tanh\left(  a\eta\right)  .\ $A Lorentz
transformation can be made to the instantaneous rest frame of the entire POVM
and in this reference frame $\eta=0$. A time interval on $\eta=0$ is then a
proper time interval $d\tau$ and from (\ref{RinderCoordinates}) $d\tau
=\exp\left(  a\xi\right)  d\eta$ in wedge I while from
(\ref{RindlerCoordinatesII}) $d\tau=-\exp\left(  a\xi\right)  d\eta$ in wedge
II. The acceleration $d^{2}x/dt^{2}$ evaluated in the rest frame of the POVM
equals $\alpha$ in wedge I and $-\alpha$ in wedge II where
\begin{equation}
\alpha\equiv a\exp\left(  -a\xi\right)  . \label{ProperAcceleraton}%
\end{equation}

With the above definitions the limit $a\rightarrow0$ is undefined in
(\ref{InOut}). To avoid this problem a translated Minkowski spatial coordinate%
\begin{equation}
X=x-a^{-1}. \label{Shifted}%
\end{equation}
can be defined and substituted in (\ref{RinderCoordinates}) to (\ref{InOut}).
In terms of $X$ the infinite $\alpha$ singularity is shifted to $X=-a^{-1}$,
but for any fixed $a$ it still exists since here are $X$ coordinates beyond
it. If the $a\rightarrow0$ limit is taken first (\ref{RinderCoordinates})
gives $t=\eta$ and $X=\xi$ in wedge I, while (\ref{RindlerCoordinatesII})
gives $X=-\infty$ in all of wedge II and the infinite $\alpha$ singularity is eliminated.

\section{Photon counting}

In this Section the detector model used here is described and the derivation
of the inertial photon position POVM is reviewed. The thickness and band
structure of the photon counting device should be matched to the frequencies
present in the incident photon field. The device should be thick enough to
absorb all photons incident on its surface. To count photons arriving from the
past the photon counting device should be prepared in its ground state, but
the basis includes negative frequency states and if the device is prepared in
an excited state it will emit.

The derivations in \cite{HawtonPOVM} and \cite{HawtonPhotonLocation} are
summarized in the next two paragraphs as follows: According to Glauber's
photodetection theory \cite{Glauber} the probability for an atom to absorb a
photon summed over the unobserved final state is proportional to $\left\langle
\widehat{E}^{2}\right\rangle =\left\langle \psi\left\vert \widehat{E}^{\left(
+\right)  \dagger}\left(  t,\mathbf{x}\right)  \widehat{E}^{\left(  +\right)
}\left(  t,\mathbf{x}\right)  \right\vert \psi\right\rangle $ where
$\widehat{E}^{\left(  +\right)  }$ is the positive frequency electric field
operator and $\left\vert \psi\right\rangle $ is the state vector of the
electromagnetic field. This absorption probability per atom is proportional to
$\omega$ or energy, not photon number. If the detector is a semiconducting
pn-junction the electric field separates the electron-hole pair created by the
photon so emission can be neglected in an ideal device. Inside the detector
the field decays as $\exp\left(  -\alpha_{\omega}x\right)  $ where
the\ absorptivity $\alpha_{\omega}$ is proportional to $\omega$.\ The positive
$x$-axis can be defined parallel to the inward normal with $y$ and $z$ in the
plane of the detector surface at $x=x^{\prime}$. For a field with definite
frequency the integral over thickness of $\left\langle \widehat{E}%
^{2}\right\rangle \propto\int_{x^{\prime}}^{\infty}dxe^{-2\alpha_{w}\left(
x-x^{\prime}\right)  }=\omega/2\alpha_{\omega}$ is frequency independent so
the detector counts photons. For a spectrally narrow single photon pulse with
definite polarization it was proved in \cite{HawtonPOVM} that the probability
density to absorb a photon in a particular pixel is the integral of
$\left\vert \left\langle u_{\mathbf{x},M}|\psi\right\rangle \right\vert ^{2}$
over pixel area and measurement time. For a spatially localized pulse that is
spectrally wide the penetration depth is not well d%
\begin{equation}
n^{\left(  +\right)  }\left(  \mathbf{x,}t-\tau\right)  =\left\vert
\left\langle u_{\mathbf{x},M}|\psi\left(  t-\tau\right)  \right\rangle
\right\vert ^{2} \label{Delay}%
\end{equation}
where $\left\langle u_{\mathbf{x},M}|\psi\right\rangle $ is the projection of
$\left\vert \psi\right\rangle $ onto the basis of exactly localized states
$\left\vert u_{\mathbf{x},M}\right\rangle $ and $\tau$ is a delay of a few
optical cycles required for the photon to be absorbed. The probability to
absorb a photon in a particular pixel is the integral of $\left\vert
\left\langle u_{\mathbf{x},M}|\psi\right\rangle \right\vert ^{2}$ over pixel
area and measurement time. For a spatially localized pulse that is spectrally
wide the penetration depth is not well defined due to the nonlocal
relationship between photon number density and the field.

In 2D a photon counting array degenerates to a single pixel that can record
arrival time \cite{HawtonPhotonLocation}. The worldline of a single photon
counting device at rest relative to the observer is sketched as the vertical
gray band in Fig.1a. In QFT particles are counted on a spacelike Cauchy
surface relative to which positive and negative frequencies can be identified
and annihilation and creation operators defined. A detector array described by
a POVM whose elements are projectors onto the localized states on the
spacelike hypersurface $t=t^{\prime}$ is sketched as the horizontal gray boxes
in Fig. 1a. The POVM has a\ timelike normal in the direction of increasing
time so it is spacelike. A moving device and detector array, both with
velocity $\beta$ relative to the observer, are sketched in Fig. 1b. The single
photon counting device travels on the worldline $x=\beta t$ but the
hypersurface $t=\beta x$ on which the POVM resides is not a world line;
instead $t$ is the time when a photon enters the hyperpixel in which it will
be counted.%
\begin{figure}
[ptb]
\begin{center}
\includegraphics[
height=3.2232in,
width=3.0398in
]%
{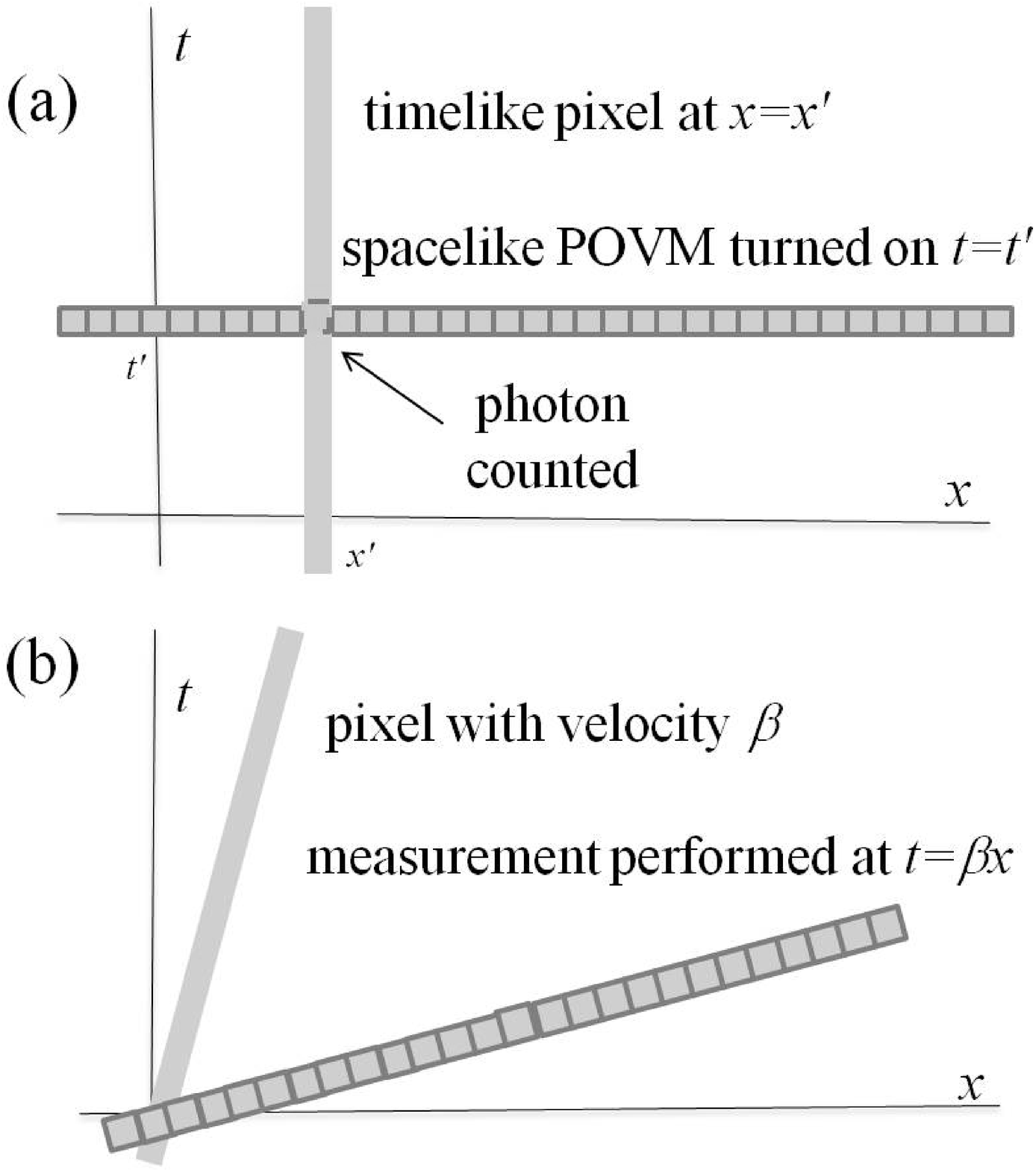}%
\caption{Inertial detectors. (a) Stationary spacelike POVM on $t=t^{\prime}$
and worldline of the pixel at $x=x^{\prime}$. The pixel at $x=x^{\prime}$
absorbs a single photon in the sketch but the array can count $0,\ 1,\ 2,$ ..
photons. (b) Moving spacelike POVM with velocity $\beta$ on the hypersurface
$t=\beta x$ and worldline $x=\beta t$ of one of its hyperpixels.}%
\end{center}
\end{figure}

A spacelike POVM describing an array of accelerated detectors is sketched in
Fig. 2. According to (\ref{ProperAcceleraton}) each point on the array has a
different acceleration $\alpha.$ This guarantees that the array is rigid in
the sense that its "$3$-geometry as seen from its own momentary rest frame is
unchanged in the course of proper time" \cite{MMdelRay}. The acceleration will
dependent on penetration depth into the semiconductor but it is assumed that
the photon is absorbed so these details are ignored. The solid and dashed
lines in Fig. 2 represent surfaces of single accelerated devices traveling on
the hyperbolic worldlines $x^{2}-t^{2}=a^{-2}\exp\left(  2a\xi\right)  $ in
wedges I and II.
\begin{figure}
[ptb]
\begin{center}
\includegraphics[
height=2.5633in,
width=3.4039in
]%
{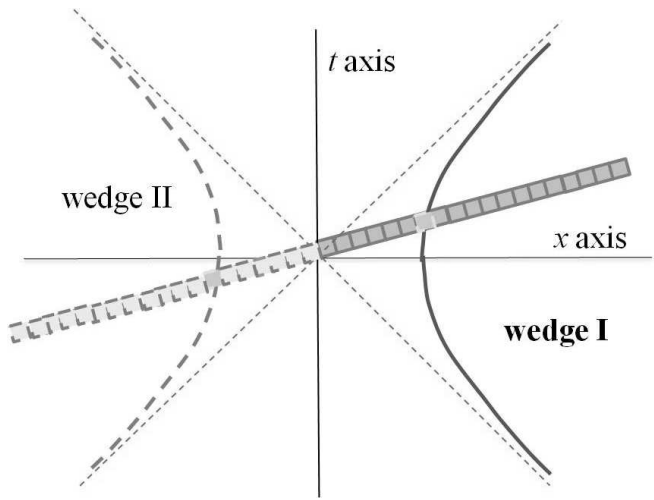}%
\caption{Rindler spacelike POVM on $\eta=\eta^{\prime}$ (gray boxes). The
solid (dashed) line is the worldline of one of its hyperpixels in wedge I
(II).}%
\end{center}
\end{figure}

\section{Photon number density and scalar product}

This section starts with a review of the 4D photon counting operator. The
Minkowski vacuum appears thermal to an accelerated observer and operators are
useful for describing absorption and emission in this multiphoton state. For a
paraxial beam with definite polarization and negligible transverse wave
vectors the problem is effectively two dimensional and the 2D approximation
will be used is the remainder of this paper for simplicity. In 2D a definite
polarization photon field is mathematically equivalent to the zero mass Klein
Gordon field. While the 2D approximation is problematic because of infrared
divergences \cite{Coleman}, it is a good model for explaining the Unruh effect
\cite{Crispino}.

The invariance of the indefinite scalar product follows from the continuity
equation for four-current \cite{Hawking}. For photons the four-current
operator can be written in terms of the four-vector potential operator as
follows \cite{HawtonMelde}: In the Heisenberg picture the positive frequency
part of the four-potential operator in the Lorentz gauge is%
\begin{equation}
\widehat{A}^{\left(  +\right)  \mu}\left(  t,\mathbf{x}\right)  =\sum
_{\lambda}\int_{-\infty}^{\infty}d^{3}k\frac{\exp\left(  i\mathbf{k\cdot
x}-i\left\vert \mathbf{k}\right\vert t\right)  }{2\pi\left(  4\pi\left\vert
\mathbf{k}\right\vert \right)  ^{1/2}}e_{\lambda}^{\mu}\left(  \mathbf{k}%
\right)  \widehat{a}_{\lambda}\left(  \mathbf{k}\right)
\label{PotentialOperator}%
\end{equation}
where $\widehat{a}_{\lambda}\left(  \mathbf{k}\right)  $ annihilates a photon
with wave vector $\mathbf{k\ }$and polarization $\lambda$. Its negative
frequency part is $\widehat{A}^{\left(  -\right)  \mu}=\widehat{A}^{\left(
+\right)  \mu\dagger}$. Contraction of (\ref{PotentialOperator}) with the
second rank electromagnetic force tensor operator%
\begin{equation}
\widehat{F}^{\left(  +\right)  \mu\nu}=\partial^{\mu}\widehat{A}^{\left(
+\right)  \nu}-\partial^{\nu}\widehat{A}^{\left(  +\right)  \mu} \label{EB}%
\end{equation}
gives the covariant four-flux operator
\begin{equation}
\widehat{J}^{\mu}=i\widehat{F}^{\left(  -\right)  \mu\nu}\widehat{A}_{\nu
}^{\left(  +\right)  }-i\widehat{A}_{\nu}^{\left(  -\right)  }\widehat
{F}^{\left(  +\right)  \mu\nu} \label{Current}%
\end{equation}
that satisfies a continuity equation. Its $0^{th}$ component is the number
density operator
\begin{equation}
\widehat{J}^{0}=i\widehat{\mathbf{E}}^{\left(  -\right)  }\cdot\widehat
{\mathbf{A}}^{\left(  +\right)  }-i\widehat{\mathbf{A}}^{\left(  -\right)
}\cdot\widehat{\mathbf{E}}^{\left(  +\right)  }. \label{nOperator}%
\end{equation}
A field is a quantity defined at every point of spacetime so, mathematically,
both $\widehat{A}$ and $\widehat{F}$ are field operators. Potential operators
can be identified by their $\left\vert \mathbf{k}\right\vert ^{-1/2}$
dependance while $\widehat{F}$ whose elements are components of the electric
and magnetic field operators has terms proportional to $\left\vert
\mathbf{k}\right\vert ^{1/2}$. \ The extra factor $\left\vert \mathbf{k}%
\right\vert $ comes from the time derivative in $\mathbf{E}=-\partial
\mathbf{A}/\partial t$. For a plane wave with definite frequency the factor
$\left\vert \mathbf{k}\right\vert ^{1/2}$ from $\mathbf{E}$ compensates for
the factor $\left\vert \mathbf{k}\right\vert ^{-1/2}$ from $\mathbf{A}$ to
give the frequency independent probability that characterizes number density.

Once $\Sigma$ has been selected a transformation can be made to the Coulomb
gauge where photons are described by the transverse vector potential alone. In
2D $\widehat{A}^{\mu}$ then has a single component which for consistency with
QFT will be called $\widehat{\phi}$. The electric field operator is
$\widehat{E}=-\partial\widehat{\phi}/\partial t$. In a Minkowski plane wave
expansion the positive frequency part of $\widehat{\phi}$ is%
\begin{equation}
\widehat{\phi}^{\left(  +\right)  }\left(  t,x\right)  =\int_{-\infty}%
^{\infty}dku_{k,M}\left(  t,x\right)  \widehat{a}_{k,M} \label{FieldOperator}%
\end{equation}
where $u_{k,M}$ are the usual Minkowski plane waves given here by
(\ref{MinkowskiPlaneWaves}). The absorption density operator (\ref{nOperator})
then reduces to%
\begin{equation}
\widehat{n}^{\left(  +\right)  }\left(  t,x\right)  =i\widehat{\phi}^{\left(
-\right)  }\overleftrightarrow{\partial}_{t}\widehat{\phi}^{\left(  +\right)
} \label{n2D+}%
\end{equation}
where $f\overleftrightarrow{\partial}_{t}g\equiv f\partial_{t}g-\left(
\partial_{t}f\right)  g$, while the emission density operator is%
\begin{equation}
\widehat{n}^{\left(  -\right)  }\left(  t,x\right)  =i\widehat{\phi}^{\left(
+\right)  }\overleftrightarrow{\partial}_{t}\widehat{\phi}^{\left(  -\right)
}. \label{n2D-}%
\end{equation}
For an electromagnetic field initially in the state $\left\vert \psi
\right\rangle $ the probability density to count a photon at $\left(
t^{\prime},x^{\prime}\right)  $ is \cite{HawtonPOVM}%
\begin{equation}
w_{1}^{\left(  +\right)  }\left(  t^{\prime},x^{\prime}\right)  =\left\langle
\psi\left\vert \widehat{n}^{\left(  +\right)  }\left(  t^{\prime},x^{\prime
}\right)  \right\vert \psi\right\rangle . \label{PhotonDensity}%
\end{equation}
For $\left(  t^{\prime},x^{\prime}\right)  $ and $\left(  t^{\prime\prime
},x^{\prime\prime}\right)  $ in different pixels the field operators commute
and the two photon correlation function can be written as
\begin{equation}
w_{2}^{\left(  +\right)  }=\left\langle \psi\left\vert \widehat{n}^{\left(
+\right)  }\left(  t^{\prime},x^{\prime}\right)  \widehat{n}^{\left(
+\right)  }\left(  t^{\prime\prime},x^{\prime\prime}\right)  \right\vert
\psi\right\rangle . \label{CoincidenceRate}%
\end{equation}

To make a connection between the number density operator and the invariant
scalar product positive and negative frequency fields can be defined for a one
photon state $\left\vert \psi\right\rangle $ as%
\begin{align}
\psi^{\left(  +\right)  }\left(  t,x\right)   &  =\int_{-\infty}^{\infty
}dk\left\langle 0\left\vert \widehat{a}_{k}\right\vert \psi\right\rangle
u_{k,M}\left(  t,x\right)  ,\label{+Field}\\
\psi^{\left(  -\right)  }\left(  t,x\right)   &  =\int_{-\infty}^{\infty
}dk\left\langle \psi\left\vert \widehat{a}_{k}^{\dagger}\right\vert
0\right\rangle u_{k,M}^{\ast}\left(  t,x\right)  ,\label{-Field}\\
\psi\left(  t,x\right)   &  =\psi^{\left(  +\right)  }\left(  t,x\right)
+\psi^{\left(  -\right)  }\left(  t,x\right)  . \label{Field}%
\end{align}
These fields are potential-like since they contain a factor $\left\vert
\mathbf{k}\right\vert ^{-1/2}$. The invariant indefinite scalar product
evaluated on the hypersurface $t=t^{\prime}$ is%
\begin{align}
\left(  \chi,\psi\right)   &  =\int_{-\infty}^{\infty}dx\left[  \left\langle
\chi\left\vert \widehat{n}^{\left(  +\right)  }\left(  t,x\right)  \right\vert
\psi\right\rangle +\left\langle \chi\left\vert \widehat{n}^{\left(  -\right)
}\left(  t,x\right)  \right\vert \psi\right\rangle \right] \nonumber\\
&  =i\int_{-\infty}^{\infty}dx\chi^{\ast}\left(  t,x\right)
\overleftrightarrow{\partial}_{t}\psi\left(  t,x\right)  .
\label{xInnerProduct}%
\end{align}
Integration over $x$ using $\left(  u_{k^{\prime\prime},M},u_{k^{\prime}%
,M}\right)  =\delta\left(  k^{\prime}-k^{\prime\prime}\right)  $ gives the
$k$-space form of the scalar product
\begin{align}
\left(  \chi,\psi\right)   &  =\int_{-\infty}^{\infty}\frac{dk}{2\left\vert
k\right\vert }\left[  \chi^{\left(  +\right)  \ast}\left(  k\right)
\psi^{\left(  +\right)  }\left(  k\right)  \right. \nonumber\\
&  \left.  -\chi^{\left(  -\right)  \ast}\left(  k\right)  \psi^{\left(
-\right)  }\left(  k\right)  \right]  . \label{kInnerProduct}%
\end{align}
Analogous expressions to (\ref{FieldOperator}) to (\ref{kInnerProduct}) exist
for wedge I and II Rindler coordinates.

\section{Plane waves}

The positive frequency Minkowski plane waves are \cite{BirrellDavies}
\begin{equation}
u_{k^{\prime},M}\left(  t,x\right)  =\frac{\exp\left[  ik^{\prime}\left(
x-\epsilon_{k^{\prime}}t\right)  \right]  }{\sqrt{4\pi\left\vert k^{\prime
}\right\vert }} \label{MinkowskiPlaneWaves}%
\end{equation}
where $\epsilon_{k}=k/\left\vert k\right\vert $ is the sign of the Minkowski
wave vector so the frequency $\epsilon_{k}k$ is positive. The factor $\left(
2\left\vert k^{\prime}\right\vert \right)  ^{1/2}$ in the denominator is
needed to compensate for $\overleftrightarrow{\partial}_{t}$ in the invariant
scalar product. Only positive frequency basis states will be written down
explicitly but the negative frequency states are just their complex
conjugates. The prime denotes a definite value while unprimed coordinates are
variable. On any $t=t^{\prime}$ hypersurface substitution of
(\ref{MinkowskiPlaneWaves}) in (\ref{xInnerProduct}) gives the indefinite
orthonormality relations%
\begin{align}
\left(  u_{k^{\prime\prime},M},u_{k^{\prime},M}\right)   &  =\delta\left(
k^{\prime}-k^{\prime\prime}\right)  ,\label{Orthonormal}\\
\left(  u_{k^{\prime\prime},M}^{\ast},u_{k^{\prime},M}^{\ast}\right)   &
=-\delta\left(  k^{\prime}-k^{\prime\prime}\right)  ,\nonumber\\
\left(  u_{k^{\prime\prime},M}^{\ast},u_{k^{\prime},M}\right)   &  =\left(
u_{k^{\prime\prime},M},u_{k^{\prime},M}^{\ast}\right)  =0.\nonumber
\end{align}
In $k$-space the Minkowski plane wave with wave vector $k^{\prime}$ is%
\begin{equation}
u_{k^{\prime},M}\left(  k\right)  =\sqrt{2\left\vert k\right\vert }%
\delta\left(  k-k^{\prime}\right)  . \label{PlaneWaveInkSpace}%
\end{equation}
It differs from the probability amplitude $\left(  u_{k,M},u_{k^{\prime}%
,M}\right)  $ by the factor $\sqrt{2\left\vert k\right\vert }$ needed to give
the orthonormality relations (\ref{Orthonormal}) with the scalar product
(\ref{kInnerProduct}).

The positive frequency Rindler plane waves are%
\begin{align}
u_{K^{\prime},I}\left(  \eta,\xi\right)   &  =\frac{\exp\left[  iK^{\prime
}\left(  \xi-\epsilon_{K^{\prime}}\eta\right)  \right]  }{\sqrt{4\pi\left\vert
K^{\prime}\right\vert }}\text{ in I}\label{RindlerPlaneWaves}\\
&  =0\text{ in II}\nonumber\\
u_{K^{\prime},II}\left(  \eta,\xi\right)   &  =\frac{\exp\left[  iK^{\prime
}\left(  \xi+\epsilon_{K^{\prime}}\eta\right)  \right]  }{\sqrt{4\pi\left\vert
K^{\prime}\right\vert }}\text{ in II}\nonumber\\
&  =0\text{ in I}\nonumber
\end{align}
where $\epsilon_{K}=K/\left\vert K\right\vert $. Positive frequency is defined
from the perspective of an inertial observer. The sign of the coefficient of
$\eta$ is reversed in wedge II because according to
(\ref{RindlerCoordinatesII}) an inertial observer sees increasing Rindler time
$\eta$ as decreasing Minkowski time $t$. This is the sign convention used in
\cite{BirrellDavies,Carroll}. In either wedge an inertial observer sees
$K^{\prime}>0$ as outward propagation and $K^{\prime}<0$ as inward propagation
in Fig. 2. On $\eta=\eta^{\prime}$ the Rindler plane waves in either wedge
satisfy orthonormality relations analogous to (\ref{Orthonormal}). Rindler
plane waves in different wedges are orthogonal.

The Minkowski and Rindler plane waves are related though the Bogoliubov
coefficients \cite{BirrellDavies} that will be evaluated by substituting
(\ref{MinkowskiPlaneWaves}) and (\ref{RindlerPlaneWaves}) in
(\ref{xInnerProduct}) and then substituting (\ref{InOut}). This direct
calculation that is the 2D version of \cite{Longhi} was performed to better
understand the relationship between the in and out waves and positive and
negative frequencies, all of which are needed here to define the localized
states. Since the scalar product is invariant $\left(  u_{k,M},u_{K,I}\right)
$ can be evaluated on any hypersurface. On $t^{\prime}=\eta^{\prime}=0$ where
the Rindler POVM is instantaneously at rest%

\begin{align}
\alpha_{kK}^{I}  &  =\left(  u_{k,M},u_{K,I}\right) \label{kRtoM}\\
&  =\int_{x_{0}}^{\infty}dx\frac{\left\vert K\right\vert /\left(  ax\right)
+\left\vert k\right\vert }{4\pi\sqrt{\left\vert kK\right\vert }}%
e^{-ikx}\left(  ax\right)  ^{iK/a},\nonumber\\
\beta_{kK}^{I}  &  =\left(  u_{k,M}^{\ast},u_{K,I}\right) \nonumber\\
&  =\int_{x_{0}}^{\infty}dx\frac{\left\vert K\right\vert /\left(  ax\right)
-\left\vert k\right\vert }{4\pi\sqrt{\left\vert kK\right\vert }}e^{ikx}\left(
ax\right)  ^{iK/a}.\nonumber
\end{align}
\ The lower limit $x_{0}$ was introduced because the $\left\vert K\right\vert
/x$ integral is undefined at $x=0.$ Using Mathematica to evaluate these
integrals,
\begin{align}
\alpha_{kK}^{I}  &  =\frac{\sqrt{\left\vert K\right\vert }}{2\pi
a\sqrt{\left\vert k\right\vert }}\left(  \frac{\left\vert k\right\vert }%
{a}\right)  ^{-iK/a}\Gamma\left(  iK/a\right)  e^{\pi\left\vert K\right\vert
/2a}\delta_{\epsilon_{K},\epsilon_{k}}\label{alphakK}\\
&  +F\left(  K\right)  \left(  ax_{0}\right)  ^{iK},\nonumber\\
\beta_{kK}^{I}  &  =-\frac{\sqrt{\left\vert K\right\vert }}{2\pi
a\sqrt{\left\vert k\right\vert }}\left(  \frac{\left\vert k\right\vert }%
{a}\right)  ^{-iK/a}\Gamma\left(  iK/a\right)  e^{-\pi\left\vert K\right\vert
/2a}\delta_{\epsilon_{K},\epsilon_{k}}\nonumber\\
&  +F\left(  K\right)  \left(  ax_{0}\right)  ^{iK}.\nonumber
\end{align}
The factor $\delta_{\epsilon_{K},\epsilon_{k}}$ excludes antiparallel
Minkowski and wedge I Rindler wave vectors. $\alpha_{kK}^{I}$ $\left(
\beta_{kK}^{I}\right)  $ is the amplitude for the positive (negative)
frequency Minkowski plane wave components of a positive frequency Rindler
plane wave in wedge I. The rest of the Bogoliubov coefficients in wedge I can
be found using
\begin{equation}
\left(  \phi,\psi\right)  =-\left(  \phi^{\ast},\psi^{\ast}\right)  =\left(
\psi,\phi\right)  ^{\ast}. \label{Others}%
\end{equation}
For negative frequency Rindler plane waves $\left(  u_{k,M}^{\ast}%
,u_{K,J}^{\ast}\right)  =-\alpha_{kK}^{I}$ and $\left(  u_{k,M},u_{K,J}^{\ast
}\right)  =-\beta_{kK}^{I}.$ If the positive frequency Minkowski plane waves
are expanded in the Rindler basis the Bololiubov coefficients are $\left(
u_{K,I},u_{k,M}\right)  =\alpha_{kK}^{I\ast}$ and $\left(  u_{K,I}^{\ast
},u_{k,M}\right)  =-\beta_{kK}^{I\ast}$. In wedge II, on $t=0$, $i\partial
_{t}u_{K,II}=\left(  \Omega/\left\vert x\right\vert \right)  u_{K,II}$ and
$u_{K,II}=\left\vert ax\right\vert ^{iK}/\sqrt{4\pi\left\vert K\right\vert }$
so the Bogoliubov coefficients are of the form (\ref{kRtoM}) but with
$x\rightarrow\left\vert x\right\vert $. Evaluation of $\alpha_{kK}^{II}$ and
$\beta_{kK}^{II}$ using Mathematica gives
\begin{equation}
\delta_{\epsilon_{K},\epsilon_{k}}\text{ in I}\rightarrow-\delta
_{\epsilon_{-K},\epsilon_{k}}\text{ in II} \label{IIfromI}%
\end{equation}
in (\ref{alphakK}).\ 

In the Minkowski vacuum an inertial device can only emit but a negative
frequency Minkowski plane wave describing emission has both positive and
negative frequency Rindler plane wave components so an accelerated device can
absorb photons from the Minkowski vacuum. The probability density for
absorption is
\begin{equation}
\left\vert \beta_{kK}^{J}\right\vert ^{2}=\frac{1}{2\pi a\left\vert
k\right\vert \left(  e^{2\pi\left\vert K\right\vert /a}-1\right)  }.
\label{kThermal}%
\end{equation}
The Minkowski vacuum is thermal with Unruh temperature $T_{U}=a/2\pi$. The
state vector describing the Minkowski vacuum in Rindler coordinates can be
deduced from the Bogoliubov transformation \cite{Crispino}. Since
(\ref{alphakK}) to (\ref{IIfromI}) give $\left(  u_{k,M}^{\ast},u_{K,I}%
+e^{-\pi\left\vert K\right\vert /a}u_{-K,II}^{\ast}\right)  =0$ the normalized
Unruh modes
\begin{align}
U_{K,I}  &  =\frac{u_{K,I}+e^{-\pi\left\vert K\right\vert /a}u_{-K,II}^{\ast}%
}{\sqrt{1-e^{-2\pi\left\vert K\right\vert /a}}},\label{UnruhStates}\\
U_{K,II}  &  =\frac{u_{K,II}+e^{-\pi\left\vert K\right\vert /a}u_{-K,I}^{\ast
}}{\sqrt{1-e^{-2\pi\left\vert K\right\vert /a}}},\nonumber
\end{align}
are purely positive frequency Minkowski states. The field operator expanded in
Rindler plane waves is%
\begin{equation}
\widehat{\phi}=\sum_{J=I,II}\int_{-\infty}^{\infty}dK\left(  u_{K,J}%
\widehat{b}_{K,J}+u_{K,J}^{\ast}\widehat{b}_{K,J}^{\dagger}\right)
\label{Rindler FieldOperator}%
\end{equation}
where $\widehat{b}_{K,J}$ annihilates a photon with wave vector $K$ in wedge
$J$. When transformed to Unruh modes
\begin{align}
\widehat{\phi}  &  =\int_{-\infty}^{\infty}dK\left[  U_{K,I}\left(
\frac{\widehat{b}_{K,I}-e^{-\pi\left\vert K\right\vert /a}\widehat{b}%
_{-K,II}^{\dagger}}{\sqrt{1-e^{-2\pi\left\vert K\right\vert /a}}}\right)
\right. \label{UnruhFieldOperator}\\
&  \left.  +U_{K,II}\left(  \frac{\widehat{b}_{K,II}-e^{-\pi\left\vert
K\right\vert /a}\widehat{b}_{-K,I}^{\dagger}}{\sqrt{1-e^{-2\pi\left\vert
K\right\vert /a}}}\right)  +H.c.\right] \nonumber
\end{align}
where the coefficients of $U_{K,J}$ annihilate the Minkowski vacuum. To write
an expression for the Minkowski vacuum state in Rindler coordinates the wave
vectors can be made discrete using periodic boundary conditions on $x=-L/2$
and $L/2$ where $K_{j}=j2\pi/L$ and then taking the limit as $L\rightarrow
\infty$ to regain the continuum of wave vectors. For discrete wave vectors the
Minkowski vacuum in the Rindler plane wave basis is described by the state
vector%
\begin{equation}
\left\vert 0_{M}\right\rangle =\prod_{j=-\infty}^{\infty}C_{j}\sum_{n_{K_{j}%
}=0}^{\infty}e^{-n_{j}\pi\left\vert K_{j}\right\vert /a}\left\vert n_{K_{j}%
},I\right\rangle \otimes\left\vert n_{-K_{j}},II\right\rangle \label{Vacuum}%
\end{equation}
where $C_{j}=\sqrt{1-e^{-2\pi\left\vert K_{j}\right\vert /a}}$. The details
are given in \cite{Crispino}. Eq. (\ref{Vacuum}) is a product over modes of
sums over the number of correlated pairs $n_{j}$.

\section{Localized states}

The plane wave (\ref{MinkowskiPlaneWaves}) has equal amplitude at all points
in space at time $t$ and phase factor $\exp\left[  ik^{\prime}\left(
x-\epsilon_{k^{\prime}}t\right)  \right]  $ while a $\delta$-function
localized state has equal amplitude for all wave vectors with a phase that
determines its position. In this section the Minkowski and Rindler plane waves
in spacetime will be converted to localized states in momentum space by
interchanging the roles of momentum and position. The recipe is
straightforward: Exchange position and wave vector by interchanging the
arguments with the subscripts. Move the $\sqrt{2\left\vert k\right\vert }$
from the denominator to the numerator to allow for the difference in the form
of the $x$-space and $k$-space scalar product. Change the sign of the $x$-term
in the exponent and interchange primed with unprimed coordinates because
$t^{\prime}$ and $x^{\prime}$ are the coordinates of a specific localization
event. With this prescription the positive frequency Minkowski localized
states on the $t=t^{\prime}$ hypersurface are%
\begin{equation}
u_{x^{\prime},M}\left(  k\right)  =\frac{\sqrt{2\left\vert k\right\vert }%
\exp\left[  -ik\left(  x^{\prime}-\epsilon_{k}t^{\prime}\right)  \right]
}{\sqrt{2\pi}} \label{MinkowskiLocalized}%
\end{equation}
for $-\infty>k>\infty$. The probability amplitude for wave vector $k$ is
\begin{equation}
\left(  u_{k,M},u_{x^{\prime},M}\right)  =\frac{\exp\left[  -ik\left(
x^{\prime}-\epsilon_{k}t^{\prime}\right)  \right]  }{\sqrt{2\pi}}.
\label{ProbAmpLocalized}%
\end{equation}
Since $\widehat{a}_{k}\left\vert 0\right\rangle =\left\vert u_{k,M}%
\right\rangle $, $\left(  u_{k,M},u_{x^{\prime},M}\right)  =\left\langle
0\left\vert \widehat{a}_{k}\right\vert u_{x^{\prime},M}\right\rangle $ so the
spacetime the field (\ref{+Field}) describing the localized state at $\left(
t^{\prime},x^{\prime}\right)  $ is
\begin{equation}
u_{x^{\prime},M}\left(  t,x\right)  =\int_{-\infty}^{\infty}dk\frac
{\exp\left[  ik\left(  x-x^{\prime}\right)  -i\epsilon_{k}k\left(
t-t^{\prime}\right)  \right]  }{2\pi\sqrt{2\left\vert k\right\vert }}.
\label{SpaceTimeLocalized}%
\end{equation}
At $t=t^{\prime}$ this can be integrated to give\ $u_{x^{\prime},M}=\sqrt
{2\pi}\left\vert x-x^{\prime}\right\vert ^{-1/2}$ which is clearly nonlocal.

The Minkowski localized states (\ref{MinkowskiLocalized}) are orthonormal.
This can be verified by substitution in the $k$-space scalar product
(\ref{kInnerProduct}) to give%
\begin{align}
\left(  u_{x^{\prime},M},u_{x^{\prime\prime},M}\right)   &  =\delta\left(
x^{\prime}-x^{\prime\prime}\right)  ,\label{MinkowskiLocalization}\\
\left(  u_{x^{\prime},M}^{\ast},u_{x^{\prime\prime},M}^{\ast}\right)   &
=-\delta\left(  x^{\prime}-x^{\prime\prime}\right)  ,\nonumber\\
\left(  u_{x^{\prime},M},u_{x^{\prime\prime},M}^{\ast}\right)   &
=0.\nonumber
\end{align}

Following the same recipe the Rindler localized states on $\eta=\eta^{\prime}$
are%
\begin{align}
u_{\xi^{\prime},I}\left(  K\right)   &  =\frac{\sqrt{2\left\vert K\right\vert
}\exp\left[  -iK\left(  \xi^{\prime}-\epsilon_{K}\eta^{\prime}\right)
\right]  }{\sqrt{2\pi}}\text{ in I}\label{RindlerLocalizedStates}\\
&  =0\text{ in II}\nonumber\\
u_{\xi^{\prime},II}\left(  K\right)   &  =\frac{\sqrt{2\left\vert K\right\vert
}\exp\left[  -iK\left(  \xi^{\prime}+\epsilon_{K}\eta^{\prime}\right)
\right]  }{\sqrt{2\pi}}\text{ in II}\nonumber\\
&  =0\text{ in I}\nonumber
\end{align}
for $-\infty>K>\infty.$ With the definitions (\ref{RindlerLocalizedStates})
the Rindler orthonormality relations on $\eta=\eta^{\prime}$ are
\begin{align}
\left(  u_{\xi^{\prime},J},u_{\xi^{\prime\prime},J}\right)   &  =\delta\left(
\xi^{\prime}-\xi^{\prime\prime}\right)  ,\label{RindlerLocalization}\\
\left(  u_{\xi^{\prime},J}^{\ast},u_{\xi^{\prime\prime},J}^{\ast}\right)   &
=-\delta\left(  \xi^{\prime}-\xi^{\prime\prime}\right)  ,\nonumber\\
\left(  u_{\xi^{\prime},J},u_{\xi^{\prime\prime},J}^{\ast}\right)   &
=0,\nonumber\\
\left(  u_{\xi^{\prime},I},u_{\xi^{\prime\prime},II}\right)   &
=0,\ etc.\nonumber
\end{align}

\subsection{Rindler localized state as seen by a Minkowski observer}

In this subsection the probability amplitudes for the Rindler localized states
will be calculated in the Minkowski localized basis. First a Rindler plane
wave will be expanded in the Minkowski localized basis because this
intermediate result will be needed later. Then the Rindler localized states
will be transformed to the Minkowski localized basis. The transformation
coefficients from $k$ to $K$-space are $\alpha_{kK}^{I}$ and $\beta_{kK}^{I}$
given by (\ref{alphakK}). The probability amplitudes for the wedge I positive
frequency Rindler plane waves in the Minkowski localized basis are
\begin{align}
\left(  u_{x,M},u_{K,I}\right)   &  =\int_{-\infty}^{\infty}dk\frac
{\exp\left(  ikx\right)  }{\sqrt{2\pi}}\alpha_{kK}^{I},\label{uKx}\\
\left(  u_{x,M}^{\ast},u_{K,I}\right)   &  =\int_{-\infty}^{\infty}%
dk\frac{\exp\left(  -ikx\right)  }{\sqrt{2\pi}}\beta_{kK}^{I}.\nonumber
\end{align}
The $k$-integrals were evaluated analytically using Mathematica. If $x>0$%
\begin{align}
\left(  u_{x,M},u_{K,I}\right)   &  =\ \frac{\left(  ax\right)  ^{iK/a}}%
{\sqrt{2\pi ax}}e^{2\pi\left\vert K\right\vert /a-\varepsilon\left\vert
K\right\vert /a}g\left(  K\right)  +f\left(  K\right)  ,\label{uKx2}\\
\left(  u_{x,M}^{\ast},u_{K,I}\right)   &  =\frac{\left(  ax\right)  ^{iK/a}%
}{\sqrt{2\pi ax}}i\epsilon_{K}g\left(  K\right)  -f\left(  K\right)
,\nonumber
\end{align}
while if $x<0$%
\begin{align}
\left(  u_{x,M},u_{K,I}\right)   &  =-i\frac{\left\vert ax\right\vert ^{iK/a}%
}{\sqrt{2\pi ax}}\epsilon_{K}e^{\pi\left\vert K\right\vert /a}g\left(
K\right)  +f\left(  K\right)  ,\label{MinkowskiK2}\\
\left(  u_{x,M}^{\ast},u_{K,I}\right)   &  =\frac{\left\vert ax\right\vert
^{iK/a}}{\sqrt{2\pi ax}}e^{\pi\left\vert K\right\vert /a}g\left(  K\right)
-f\left(  K\right)  ,\nonumber
\end{align}
where%
\begin{align}
g\left(  K\right)   &  \equiv\frac{1}{2\pi}\left(  -1\right)  ^{1/4}%
\epsilon_{K}\sqrt{\frac{\left\vert K\right\vert }{a}}\label{g}\\
&  \times\Gamma\left(  \frac{1}{2}-i\frac{K}{a}\right)  \Gamma\left(
i\frac{K}{a}\right)  e^{-\pi\left\vert K\right\vert /a},\nonumber
\end{align}%
\begin{equation}
\left\vert g\left(  K\right)  \right\vert ^{2}=\left(  e^{4\pi\left\vert
K\right\vert /a}-1\right)  ^{-1}. \label{Thermal}%
\end{equation}
The small constant $\varepsilon$ was introduced to give a convergent integral
and a finite linewidth that makes a $\delta$-function peak visible in the
graphs. The $f\left(  K\right)  $ term is the integral of the $F\left(
K\right)  $ term in (\ref{alphakK}). For $x_{0}\ll x$
\[
f\left(  K\right)  =\frac{i\left(  ax_{0}\right)  ^{iK/a}}{\sqrt
{2\pi\left\vert K\right\vert /a}}.
\]
The factor $g\left(  K\right)  $ suggests a temperature $T=a/4\pi$ which is
half the Unruh temperature. However, the plane wave annihilation and creation
operators are the usual ones so in the Minkowski vacuum a Rindler observer
sees a thermal distribution of photon numbers characterized by the Unruh
temperature. The factor $g\left(  K\right)  $ is the result of integration
over $k$ taking into account the nonlocality of the fields. For $x>0$ the
emission and absorption probabilities look like $T_{U}/2$ but for $x<0$ the
emission and absorption probabilities are equal.

Expressions (\ref{uKx2}) and (\ref{MinkowskiK2}) are probability amplitudes
for Rindler plane waves in the Minkowski localized basis. Using these
probability amplitudes and $\left(  u_{K,I},u_{\xi,I}\right)  =\exp\left(
-iK\xi\right)  /\sqrt{2\pi}$ the transformation coefficients from wedge I
Rindler localized states to Minkowski localized states can be written as
\begin{align}
\alpha_{x\xi}^{I}  &  =\left(  u_{x,M},u_{\xi,I}\right)
\label{MinkowskiToRindler}\\
&  =\int_{-\infty}^{\infty}dK\left(  u_{x,M},u_{K,I}\right)  \frac{\exp\left(
-iK\xi\right)  }{\sqrt{2\pi}},\nonumber\\
\beta_{x\xi}^{I}  &  =\left(  u_{x,M}^{\ast},u_{\xi,I}\right) \nonumber\\
&  =\int_{-\infty}^{\infty}dK\left(  u_{x,M}^{\ast},u_{K,I}\right)  \frac
{\exp\left(  -iK\xi\right)  }{\sqrt{2\pi}}.\nonumber
\end{align}
The localized state $u_{\xi,J}$ is seen as positive frequency by a Rindler
observer while $\alpha_{x\xi}^{I}$ and $\beta_{x\xi}^{I}$ are the probability
amplitudes for positive and negative frequency Minkowski localized states
respectively. Using (\ref{Others}) the Minkowski amplitudes for the negative
frequency Rindler localized states are $\left(  u_{x,M},u_{\xi,I}^{\ast
}\right)  =-\beta_{x\xi}^{I}$ and $\left(  u_{x,M}^{\ast},u_{\xi,I}^{\ast
}\right)  =-\alpha_{x\xi}^{I}$. The inverse transformation from Minkowski
localized states to the Rindler basis is $\left(  u_{\xi,I},u_{x,M}\right)
=\alpha_{x\xi}^{I\ast}$, $\left(  u_{\xi,I}^{\ast},u_{x,M}\right)
=-\beta_{x\xi}^{I\ast}$, $\left(  u_{\xi,I},u_{x,M}^{\ast}\right)
=\beta_{x\xi}^{I\ast}$ and $\left(  u_{\xi,I}^{\ast},u_{x,M}^{\ast}\right)
=-\alpha_{x\xi}^{I\ast}$. In wedge II use of (\ref{IIfromI}) in (\ref{uKx})
gives $\alpha_{x\xi}^{II}=-\alpha_{\left\vert x\right\vert \xi}^{I}$ and
$\beta_{x\xi}^{II}=-\beta_{\left\vert x\right\vert \xi}^{I}$.

The $K$-integral in (\ref{MinkowskiToRindler}) was evaluated numerically. The
integral of the rapidly oscillating $f\left(  K\right)  $ term is zero. If the
Minkowski observer were to see the Rindler localized states $u_{\xi,J}$ as
exactly localized, all $K^{\prime}s$ should have equal weight and the graph of
$\left\vert \left(  u_{K,I},u_{x,M}\right)  \right\vert $ should be flat while
all the other $K$-integrands should be zero. Examination of Fig. 3 shows that
these conditions are fulfilled except near $K=0$ where the probability
amplitudes diverge as $\left\vert K\right\vert ^{-1/2}$. The flat region does
lead to a $\delta$-function and, without the thermal factor, $\alpha_{x\xi
}^{I}$ would equal $\delta\left(  \ln\left\vert ax\right\vert -a\xi\right)
/\left\vert ax\right\vert ^{1/2}$. The thermal peaks give an additional
delocalized component to all the curves.
\begin{figure}
[ptb]
\begin{center}
\includegraphics[
height=2.1724in,
width=3.4385in
]%
{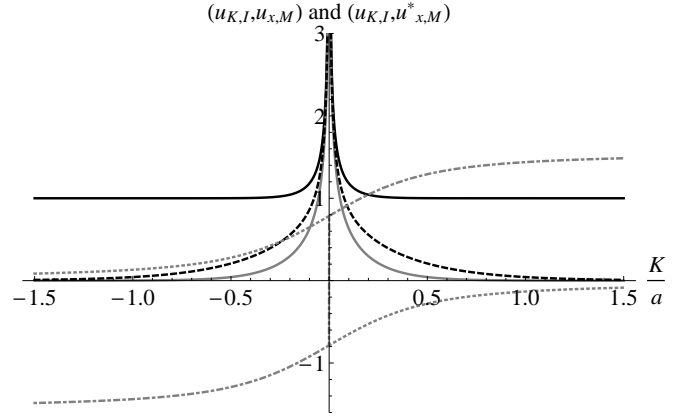}%
\caption{Thermal factors. The solid black and gray curves are plots of the
absolute value of (\ref{uKx2}) and (\ref{MinkowskiK2}) respectively for
$\left\vert ax\right\vert =1.$ The dotted and dash-dotted curves are
$\arg\left(  u_{K,I},u_{x,M}\right)  $ and $\arg\left(  u_{K,I},u_{x,M}^{\ast
}\right)  $.}%
\end{center}
\end{figure}
The real functions $\left\vert ax\right\vert ^{1/2}\alpha_{x\xi}^{I}$ and
$\left\vert ax\right\vert ^{1/2}\beta_{x\xi}^{I}$ are plotted in Fig. 4 as a
function of $a\xi-\ln\left\vert ax\right\vert $. Both the peak in
$\alpha_{x\xi}^{I}$ for $x>0$ and a delocalized contribution to all the curves
are evident. As seen by an inertial observer a positive frequency Rindler
localized state has a wedge I positive frequency peak at the correct position,
but it has additional positive and negative frequency delocalized parts that
are largest in wedge II.%

\begin{figure}
[ptb]
\begin{center}
\includegraphics[
height=2.0107in,
width=3.2733in
]%
{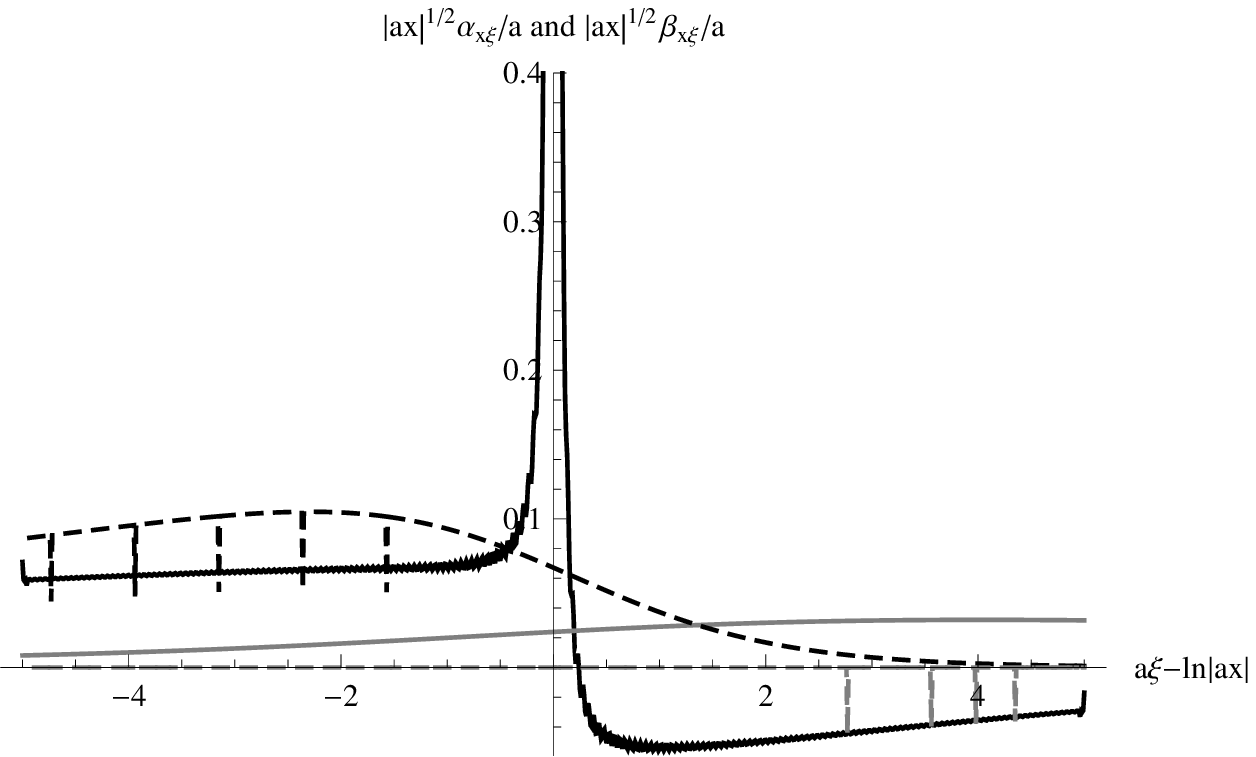}%
\caption{Real functions $\left\vert ax\right\vert ^{1/2}\alpha_{x\xi}^{I}/a$
and $\left\vert ax\right\vert ^{1/2}\beta_{x\xi}^{I}/a$ for $\varepsilon=0.01$
as a function of $a\xi-\ln\left\vert ax\right\vert $. The solid black and gray
lines are $\alpha_{x\xi}^{I}$ and $\beta_{x\xi}^{I}$ in the wedge I where
$x>0$ while the dashed line is $-\alpha_{x\xi}^{I}$ and $-\beta_{x\xi}^{I}$ in
wedge II where $x<0.$ The hash marks are individual points where the integral
did not converge.}%
\end{center}
\end{figure}

In terms of the shifted Minkowski coordinate $X=x-a^{-1}$ the horizontal axis
in Fig. 4 is%
\[
a\xi-\ln\left\vert 1+aX\right\vert \simeq a\left(  \xi-X\right)
\]
for $\left\vert X\right\vert \ll a^{-1}$. For any definite value of $a$ the
infinite $\alpha$ singularity at $X=a^{-1}$ exists and the equations and
graphs in this section apply. The order in which the $X\rightarrow\infty$ and
$a\rightarrow0$ limit is taken matters: If the limit $a\rightarrow0$ is taken
before performing the calculation Rindler plane waves reduce to Minkowski
plane waves and the graph in Fig. 4 should be replaced with $\delta$-function
at $\xi=X$.

\section{Discussion of localized states}

The definition (\ref{MinkowskiLocalized}) leading to the orthonormality
conditions (\ref{MinkowskiLocalization}) is based on Newton and Wigner's (NW)
seminal paper \cite{NW}. NW exactly localized states, and indeed any state
localized in a finite region, have paraxical properties: (i) the fields
describing localized states are themselves nonlocal, (ii) they spread
throughout space instantaneously, (iii) NW found no localized states for
photons, and (iv) negative frequency states are needed for a complete basis
and their QFT scalar products are negative. These properties are discussed
below. A one-photon state and an inertial detector are considered for simplicity.

(i) For practical purposes the most important property of a localized basis is
that the absolute square of the projection of the photon state vector onto the
the localized states is the photon number probability density on $\Sigma$,
$\left\vert \left(  u_{x,M},\psi\right)  \right\vert ^{2}$ and $\left\vert
\left(  u_{x,M}^{\ast},\psi\right)  \right\vert ^{2}$. Here $\psi$ is the
field (\ref{Field}) and $u_{x,M}$ is given by (\ref{MinkowskiLocalized}).
Propagation of $\psi$ is not the subject of this paper, but the form of
(\ref{FieldOperator}) and (\ref{MinkowskiPlaneWaves}) ensures that the field
will propagate at the speed of light. Inside the detector the damped field $E$
interacts with its atoms. The fact that a localized state is described by a
nonlocal field such as (\ref{SpaceTimeLocalized}) is not without observable
consequences in photon counting experiments: Penetration depth and hence the
time required for photon absorption are frequency dependent.

(ii) States that are localized in a finite region for an instant in time
subsequently spread instantaneously \cite{Hegerfeldt}. This causality paradox
has a straigtforward interpretation: A localized state is a sum of
counterpropagating waves whose nonlocal tails interfere destructively
\cite{Prigogine}. The exactly localized states are of this form as can be seen
by inspection of (\ref{SpaceTimeLocalized}). The field is nonlocal even on
$t=t^{\prime}$ but the spacetime probability amplitude%
\begin{equation}
\left(  u_{t,x,M},u_{x^{\prime},t^{\prime},M}\right)  =\frac{1}{2\pi}%
\int_{-\infty}^{\infty}dk\exp\left(  ik\Delta x-i\epsilon_{k}k\Delta t\right)
. \label{CounterPropagating}%
\end{equation}
with $\Delta x\equiv x-x^{\prime}$ and $\Delta t\equiv t-t^{\prime}$ is local.
By inspection (\ref{CounterPropagating}) is a sum of an $\epsilon_{k}=1$ wave
propagating to the right and an $\epsilon_{k}=-1$ wave propagating to the
left. Integration of (\ref{CounterPropagating}) using $\lim_{\varepsilon
\rightarrow0}\left(  \varepsilon\pm iz\right)  ^{-1}=\pi\delta\left(
z\right)  \mp i\mathcal{P}\left(  1/z\right)  $ where $\mathcal{P}$ is the
principal value gives%
\begin{align*}
&  \frac{1}{2}\left[  \delta\left(  \Delta x-\Delta t\right)  +\delta\left(
\Delta x+\Delta t\right)  \right] \\
&  +\frac{i}{2\pi}\left[  \mathcal{P}\left(  \frac{1}{\Delta x-\Delta
t}\right)  -\mathcal{P}\left(  \frac{1}{\Delta x+\Delta t}\right)  \right]  .
\end{align*}
For a localized state on the hypersurface $t=t^{\prime}$ the principal value
terms interfere destructively leaving $\delta\left(  x-x^{\prime}\right)  $.
At other times the nonlocal principal value terms are nonzero throughout space.

(iii) NW postulated that a localized state should be spherically symmetric
\cite{NW}. It can be proved that there is no photon position operator with
commuting components that transforms like a vector \cite{Jordan78}. Because
photon spin and orbital angular momentum cannot be separated, their localized
states have definite total angular momentum and a vortex structure like
twisted light \cite{AllenBarnettPadgett}. When this is taken into account
localized photon states and a photon position operator with commuting
components can be constructed using the NW procedure \cite{HawtonBaylis}. This
photon position operator does not transform like a vector due to the extra
term required to rotate its axis of symmetry, so the nonexistence proofs do
not apply to it.

(iv) The indefinite scalar product (\ref{xInnerProduct}) is positive for
positive frequency basis states and negative for negative frequency states.
Since positive frequency is associated with absorption and negative frequency
with emission, the integral over $\Sigma$ gives the number of photons absorbed
minus the number emitted. What matters is net absorption and a photon that is
reemitted is not counted. This is reasonable since an atomic transition can be
accompanied by an aborbed or emitted photon. Integration over time is needed
to separate these processes and inertial and accelerated observers experience
a different proper time. The invariance of the scalar product tells us that
inertial and accelerated observers will agree on the number of atomic
transitions and the probability of absorption minus the probability of
emission but they will not agree on whether photons are emitted or absorbed.

Finally, the sum over histories gives a novel path-integral interpretation of
the NW localized states that supports the interpretation that they describe a
photon counting experiment: If the NW propagator is used in a relativistic
calculation of the sum over paths crossing a intermediate spacelike
hypersurface $\Sigma$ these paths must terminate on $\Sigma$
\cite{HalliwellOrtiz}.

\section{Applications}

In this section real devices and the Unruh effect will be discussed.
Coincidence counting, single photon absorption events, and preparation of a
one photon state will be considered.

If a device is prepared in its ground state initially it will contain no
photons along its entire length. Photons are counted when they cross one of
its surfaces travelling at speed $c$ and are absorbed within a penetration
depth of a few wavelengths. (The speed of light $c$ has been reintroduced in
this paragraph for clarity.) Calculation of the probability for this event
requires flux (photon/s) but a spacelike hypersurface is required to define
annihilation and creation operators and such a basis gives number density
(photons/m). However, at normal incidence the Minkowski flux and number
density are simply related through
\begin{equation}
\left\langle \psi\left\vert \widehat{J}^{1\left(  +\right)  }\left(
t,x\right)  \right\vert \psi\right\rangle =\pm c\left\langle \psi\left\vert
\widehat{n}^{\left(  +\right)  }\left(  t,x\right)  \right\vert \psi
\right\rangle \label{Flux}%
\end{equation}
where the flux is positive (negative) for left to right (right to left)
propagation. The localized basis states themselves do not contain information
on the propagation direction but it will be assumed that the detector counts
photons from only one direction and modes in the opposite direction are traced
out. Eq. (\ref{Flux}) gives the probability per unit time to absorb a photon
in terms of the number density calculated using a basis of positive frequency
localized states on a spacelike hypersurface.

In the vacuum state (\ref{Vacuum}) for every wedge I photon with wave vector
$K$ there is a wedge II photon with wave vector $-K$ (with the sign
conventions used here). The zero photon term does not excite the detector. To
first order in $e^{-\pi\left\vert K_{j}\right\vert /a}$ this leaves the one
pair term with $C_{j}=1$ and this approximation will be considered first.
Entanglement transfer from the vacuum to a pair of counteraccelerating
Unruh-DeWitt detectors was studied in\ \cite{Reznik}. Rindler observers in the
causally disconnected wedges I and II cannot communicate but a Minkowski
"spectator" can receive signals from accelerated detectors in both wedges.
Particle content measured by the two detectors each described by a single
localized mode is predicted to be correlated \cite{DraganBruschi}. Here a
coincidence counting experiment performed by an inertial spectator with access
to a pair of counteraccelerating photon counting detectors will be analyzed.
For detectors that can absorb photons propagating from right to left described
by the Rindler null coordinate $v\equiv\eta+\xi$ the $n_{K}=1$ term in
(\ref{Vacuum}) gives the probability amplitude for two photon absorption%
\begin{align}
\left\langle u_{v^{\prime},I}u_{v^{\prime\prime},II}|0_{M}\right\rangle  &
=\int_{-\infty}^{0}dKe^{-\pi\left\vert K\right\vert /a}\frac{e^{iK\left(
v^{\prime}-v^{\prime\prime}\right)  }}{2\pi}\nonumber\\
&  =\frac{1}{2\pi}\frac{1}{i\left(  v^{\prime}-v^{\prime\prime}\right)
+\pi/a}. \label{Correlation}%
\end{align}
This describes Lorentzian spacetime correlations with linewidth $2\pi/a$. As
$a\rightarrow0$ the linewidth becomes infinite and as $a\rightarrow\infty$
$\left\langle u_{v^{\prime},I}u_{v^{\prime\prime},II}|0_{M}\right\rangle
\rightarrow-\frac{1}{2}\delta\left(  v^{\prime}-v^{\prime\prime}\right)
-\frac{i}{2\pi}P\frac{1}{v^{\prime}-v^{\prime\prime}}$ where $P$ is the
principle value. If positive and negative $K$ are included so
(\ref{Correlation}) is a sum over counterpropagating waves $\left\langle
u_{v^{\prime},I}u_{v^{\prime\prime},II}|0_{M}\right\rangle \rightarrow
-\delta\left(  v^{\prime}-v^{\prime\prime}\right)  $ and the spacetime
correlations are exact in the infinite acceleration limit.

If the photon in wedge II is not detected the probability density to count a
photon in wedge I is the partial trace of the absolute square of
(\ref{Correlation}) over $v^{\prime\prime}$ in wedge II equal to%
\begin{equation}
\int_{-\infty}^{\infty}dv^{\prime\prime}\left\vert \left\langle u_{v^{\prime
},I}u_{v^{\prime\prime},II}|0_{M}\right\rangle \right\vert ^{2}=\frac{a}%
{4\pi^{2}}. \label{OneDetector}%
\end{equation}
Eq (\ref{OneDetector}) gives probability per unit Rindler time. The proper
time interval is $d\tau=\pm\left(  a/\alpha\right)  d\eta$ where
$\alpha=ae^{-a\xi}$ as given by (\ref{ProperAcceleraton}). The probability per
unit proper time for a detector with an absorbing surface at $\xi$ to count a
photon is thus $\alpha/4\pi^{2}$.

Now imagine a that a single accelerated detector at $\xi^{\prime}$ in wedge I
capable of absorbing photons propagating from right to left from the Minkowski
vacuum clicks at Rindler time $\eta^{\prime}$. This process is most simply
viewed from the instantaneous rest frame of the POVM where $t^{\prime}%
=\eta^{\prime}=0$. To first order the one pair term in (\ref{Vacuum}) will
collapse to a normalized one photon state described by the $K$-space and $\xi
$-space Rindler probability amplitudes
\begin{align}
\left(  u_{K,II},\psi\right)   &  =\left(  \frac{\pi}{a}\right)  ^{1/2}%
e^{-\pi\left\vert K\right\vert /a}e^{-iK\xi^{\prime}},\label{PsiCollapsed}\\
\left(  u_{\xi,II},\psi\right)   &  =\frac{2\pi}{a^{3/2}}\left[  \left(
\xi-\xi^{\prime}\right)  +i\frac{\pi}{a}\right]  ^{-1},\nonumber
\end{align}
respectively. This is analogous to preparation of a one photon state using
spontaneous parametric down conversion (SPDC). The Rindler field propagates
causally after collapse at $t=\eta=0$. At Rindler time $\eta$ the peak of the
pulse has propagated outward in wedge II to $\xi=\xi^{\prime}+\left\vert
\eta\right\vert $ where $\eta<0$ since $t>0$. This is consistent with
(\ref{Correlation}) if $\eta^{\prime}=0$ and $v=v^{\prime\prime}$ are
substituted. In the Minkowski localized basis%
\begin{equation}
\left(  u_{x,M},\psi\right)  =\int_{-\infty}^{\infty}dx\alpha_{x\xi}%
^{II}\left(  u_{\xi,II},\psi\right)  . \label{PsiCollapsedInertial}%
\end{equation}
The Minkowski field (\ref{Field}) with $\left\langle n_{k}\left\vert
\widehat{a}_{k}\right\vert \psi\right\rangle =\left(  u_{k,M},\psi\right)  $
also propagates causally after collapse. According to Fig. 4 $\alpha_{x\xi
}^{II}=-\alpha_{\left\vert x\right\vert \xi}^{I}$ has a delocalized component
that is largest in wedge II. Thus the photon state prepared when a wedge I
\ accelerated detector counts a photon can lead to absorption by an
accelerated detector in wedge II or by an inertial detector in either wedge.

The one pair term was considered first for clarity and to obtain analytic
expressions, but all $n_{j}$ in (\ref{Vacuum}) can be included by calculating
the expectation values of the Rindler one and two photon density operators
analogous to (\ref{PhotonDensity}) and (\ref{CoincidenceRate}). The Rindler
number density operators are
\begin{equation}
\widehat{n}_{J}^{\left(  +\right)  }\left(  \xi\right)  =i\widehat{\phi}%
_{J}^{\left(  -\right)  }\overleftrightarrow{\partial}_{\eta}\widehat{\phi
}_{J}^{\left(  +\right)  } \label{nJRindler}%
\end{equation}
where%
\begin{equation}
\widehat{\phi}_{J}^{\left(  +\right)  }=\int_{-\infty}^{\infty}dKu_{K,J}%
\left(  \eta,\xi\right)  \widehat{b}_{K,J} \label{PhiJ}%
\end{equation}
and $\widehat{\phi}_{J}^{\left(  -\right)  }=\widehat{\phi}_{J}^{\left(
+\right)  \dagger}$. For right to left propagation
\begin{align*}
w_{I}^{\left(  +\right)  }  &  =\left\langle 0_{M}\left\vert \widehat
{n}_{v^{\prime},I}^{\left(  +\right)  }\right\vert 0_{M}\right\rangle \\
&  =\int_{-\infty}^{0}dK\left(  e^{2\pi\left\vert K\right\vert /a}-1\right)
^{-1}%
\end{align*}
This integral diverges at $\left\vert K\right\vert =0$ but there is a lower
limit to the frequency response of any real detector. If a minimum frequency
$\Omega_{0}=\left\vert K_{0}\right\vert $ is introduced the probability
density per unit Rindler time for photon absorption is%
\begin{equation}
w_{I}^{\left(  +\right)  }=-\frac{a}{\pi}\ln\left(  1-e^{-2\pi\Omega_{0}%
/a}\right)  . \label{OneCorrelation}%
\end{equation}
In the limit $2\pi\Omega_{0}/a\ll1$ $w_{I}^{\left(  +\right)  }\cong-\left(
a/\pi\right)  \ln\left(  2\pi\Omega_{0}/a\right)  $ which diverges as
$\Omega_{0}\rightarrow0$. The two photon counting rate is
\begin{equation}
w_{I,II}^{\left(  +\right)  }=\left\langle 0_{M}\left\vert \widehat
{n}_{v^{\prime},I}\widehat{n}_{v^{\prime\prime},II}\right\vert 0_{M}%
\right\rangle . \label{TwoCorrelation}%
\end{equation}
To evaluate (\ref{TwoCorrelation}) consider the effect of annihilating a
photon pair on $\left\vert 0_{M}\right\rangle $. The $K$-factor of
$\widehat{b}_{-K,II}\widehat{b}_{K,I}\left\vert 0_{M}\right\rangle $ is
\begin{equation}
C_{K}\sum_{n_{K}=0}^{\infty}n_{K}e^{-n_{K}\pi\left\vert K\right\vert
/a}\left\vert n_{K}-1,I\right\rangle \otimes\left\vert n_{-K}%
-1,II\right\rangle . \label{Two}%
\end{equation}
The probability for the $n$-pair term of this state combined with the $n$-pair
term of $\left\vert 0_{M}\right\rangle _{K}$ is $C_{K}^{2}e^{\pi\left\vert
K\right\vert /a}n_{K}\left(  e^{-2\pi\left\vert K\right\vert /a}\right)  ^{n}%
$. Defining $x=e^{-2\pi\left\vert K\right\vert /a}$ where $\sum_{n=0}^{\infty
}nx^{n}=x/\left(  1-x\right)  ^{2}$ and $C_{K}^{2}=1-x$, the sum over $n$ in
(\ref{Two}) gives $\exp\left(  \pi\left\vert K\right\vert /a\right)  /\left[
\exp\left(  2\pi\left\vert K\right\vert /a\right)  -1\right]  $. There is also
a $\widehat{b}_{K,I}\widehat{b}_{K^{\prime},II}\left\vert 0_{M}\right\rangle $
\ term that gives the square of (\ref{OneCorrelation}). For photons
propagating from right to left
\begin{equation}
w_{I,II}^{\left(  +\right)  }=\left\vert \int_{-\infty}^{-\Omega_{0}}%
dK\frac{e^{\pi\left\vert K\right\vert /a-iK\left[  \left(  v^{\prime\prime
}-v^{\prime}\right)  \right]  }}{2\pi\left(  e^{2\pi\left\vert K\right\vert
/a}-1\right)  }\right\vert ^{2}+w_{I}^{\left(  +\right)  }w_{II}^{\left(
+\right)  }. \label{Coincidence}%
\end{equation}
The last term does not depend on the spacetime coordinates. The first term,
\begin{equation}
r_{I,II}=\left\vert \int_{-\infty}^{-\Omega_{0}}dK\frac{e^{\pi\left\vert
K\right\vert /a-iK\left(  v^{\prime\prime}-v^{\prime}\right)  }}{2\pi\left(
e^{2\pi\left\vert K\right\vert /a}-1\right)  }\right\vert ^{2},
\label{CorrelatedPart}%
\end{equation}
is plotted in Fig. 5. The dotted line at the bottom is the first order
approximation, equal to the absolute square of (\ref{Correlation}). The solid,
dashed, dotdashed, and gray curves are $\Omega_{0}/a=.01$, $.02$, $.05$, and
$.13$. The probability that an accelerated detector with cutoff frequency
$\Omega_{0}$ will absorb a photon from the Minkowski vacuum is enhanced when
all $n_{K}$ are included.
\begin{figure}
[ptb]
\begin{center}
\includegraphics[
height=1.9804in,
width=3.1782in
]%
{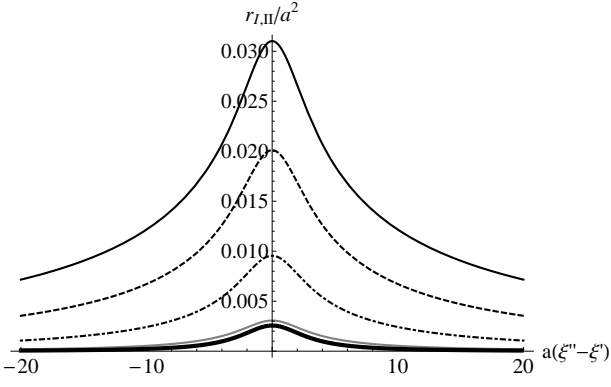}%
\caption{Coincidence counting rate$/a^{2}$ versus $a\left(  v^{\prime\prime
}-v^{\prime}\right)  $.$\ $The dotted line is the one pair term. The other
lines are $r_{I,II}^{\left(  +\right)  }$ where the solid line is $\Omega
_{0}/a=.01,$ the dashed line is $\Omega_{0}/a=.02$, the dotdashed line is
$\Omega_{0}/a=.05$ and the gray line is $\Omega_{0}/a=.13$. $\Omega_{0}$ is
the minimum (cutoff) frequency.}%
\end{center}
\end{figure}
To obtain probability densities per unit proper time the constant $a$ should
be replaced with the location dependent proper acceleration $\alpha=ae^{-a\xi
}$ as in (\ref{OneDetector}). Since $\exp\left(  \pm iKd\eta\right)
=\exp\left[  \pm i\left(  Ke^{-a\xi}\right)  d\tau\right]  $ in the
instantaneous rest frame of the detector, the minimum frequency in units of
proper time is $\omega_{0}=\Omega_{0}e^{-a\xi}$ so $\Omega_{0}/a=\omega
_{0}/\alpha$ and (counts per unit Rindler time)$/a$ equals (counts per unit
proper time)$/\alpha$ in Fig. 5 and all of the equation in this section. In SI
units $a$ can be replaced with the acceleration frequency $a/c$
\cite{Crispino} which has units $s^{-1}$ to give $\exp\left(  -2\pi\Omega
_{0}c/a\right)  $ in Eqs. (\ref{OneCorrelation}) to (\ref{CorrelatedPart}). An
Unruh temperature $T_{U}=a\hbar/2\pi ck_{B}$ of $1K$ corresponds to an
acceleration of $4\times10^{21}ms^{-2}$ and an acceleration frequency $a/c=$
$2\times10^{12}$ $s^{-1}$. For $c\Omega_{0}/a=.01$ a temperature of $1K$
requires a detector that can absorb photon with angular frequencies greater
that $2\times10^{10}$ $rad/s$.

\section{Conclusion}

In this paper bases of exactly localized Minkowski and Rindler states on
spacelike hypersurfaces $\Sigma$ were used to construct POVMs for position
measurements performed using small photon counting detectors. The
transformation coefficients from Minkowski to Rindler localized states,
$\alpha_{x\xi}^{J}$ and $\beta_{x\xi}^{J}$, were calculated and are plotted in
Fig. 4. A photon field $\psi$ was defined whose positive frequency terms
describe absorption of photons arriving from the past and whose negative
frequency terms describe emission. The absolute squares of the indefinite
scalar products of the positive frequency localized states with the field
$\left\vert \left(  u_{x,M},\psi\right)  \right\vert ^{2}$ and $\left\vert
\left(  u_{\xi,J},\psi\right)  \right\vert ^{2}$ are the probability densities
for photon absorption by inertial and accelerated devices respectively. Using
the relationship between photon density and flux this gives the probability
that a photon will enter the detector and be absorbed. The exactly localized
states are very convenient, largely because they are orthonormal and complete.
While the states defining the POVM are exactly localized, this choice of basis
does not impose limitations on the size of the photon counting detector or the
form of the field incident on it.

These photon counting POVMs were applied to vacuum excitation of accelerated
detectors (the Unruh effect). For right to left propagation described by the
null Rindler coordinate $v=\eta+\xi$ the coincidence rate for absorption of
correlated photons at $v^{\prime}$ in wedge I and $v^{\prime\prime}$ in wedge
II was found to be a Lorentzian function of $v^{\prime}-v^{\prime\prime}$ with
linewidth $2\pi/a$ where $a$ is the proper acceleration on $\xi=0$. If no
measurement is performed in wedge II the wedge I probability per unit proper
time to absorb a photon is proportional to the proper acceleration
$\alpha=a\exp\left(  -a\xi\right)  $ where $\xi$ is the Rindler coordinate of
the absorbing surface of the detector. Inclusion of numbers of photon pairs
from $1$ to $\infty$ increases the coincidence counting density by a factor
$10$ if the lowest frequency photon that can be absorbed by the detector is
$\Omega_{0}=.01a$. If a photon is absorbed by an accelerated detector in wedge
I the zero Rindler photon term is eliminated and the vacuum collapses to a one
photon state plus higher order terms. This is consistent with the conclusion
of Unruh and Wald \cite{UnruhWald}: "it seems as though the detector is
excited by swallowing part of the vacuum fluctuation of the field in the
region of spacetime containing the detector. This liberates the correlated
fluctuation in a noncausally related region of the spacetime to become a real
particle." Here liberation of a photon is interpreted as collapse, analogous
to preparation of a one photon state using SPDC. The photon state prepared in
this way can, at least in principle, lead to absorption of a photon by an
inertial detector.

\textit{Acknowledgements: }The author thanks the Natural Sciences and
Engineering Research Council for financial support.

\end{document}